\documentstyle[pre,aps,psfig,twocolumn]{revtex}
\begin{document}
\newcommand{\DB}[1]{\marginpar{\footnotesize DB: #1}}
\draft
\flushbottom
\twocolumn[
\hsize\textwidth\columnwidth\hsize\csname @twocolumnfalse\endcsname

\title{Semiclassical Inequivalence of Polygonalized Billiards} 
\author{Debabrata Biswas}
\address{Theoretical Physics Division, Bhabha Atomic Research Centre,
Trombay, Mumbai 400 085, India }

\maketitle

\tightenlines
\widetext
\advance\leftskip by 57pt
\advance\rightskip by 57pt

\begin{abstract}
Polygonalization of any smooth billiard boundary can be carried out
in several ways. We show here that the semiclassical
description depends on the polygonalization process and the
results can be inequivalent. We also establish that generalized
tangent-polygons are closest to the corresponding smooth billiard 
and for de Broglie wavelengths larger than the average length of
the edges, the two are semiclassically equivalent. 

\vskip 0.05 in
Published in Phys. Rev. E 61, 5073 (2000), \copyright The American 
Physical Society

\vskip 0.1in
\end{abstract}

]
\narrowtext
\tightenlines

\newcommand{\be}{\begin{equation}}
\newcommand{\ee}{\end{equation}}
\newcommand{\bea}{\begin{eqnarray}}
\newcommand{\eea}{\end{eqnarray}}
\newcommand{\Lop}{{\cal L}}

\section{Introduction}

Classical billiards are enclosures within which a particle moves
freely, and, on collision, reflects specularly from the boundary. 
The nature of the dynamics thus depends solely on the
shape of the enclosure and can vary from integrable motion 
in case of the circle billiard to hard chaos for the over-lapping
$3$-disk enclosure. In both these examples, the boundary consists
of smooth curves and while these are limiting cases, generic
smooth enclosures give rise to intermittent motion. As opposed
to such billiards, there exists the category of polygonal
billiards where the boundary consists of straight edges alone.
These are non-chaotic and generically non-integrable \cite{katok,gutkin}.
However, any smooth billiard can be {\em polygonalized} and in
more ways than one.

\begin{figure}[tbp]
{\hspace*{1cm}\psfig{figure=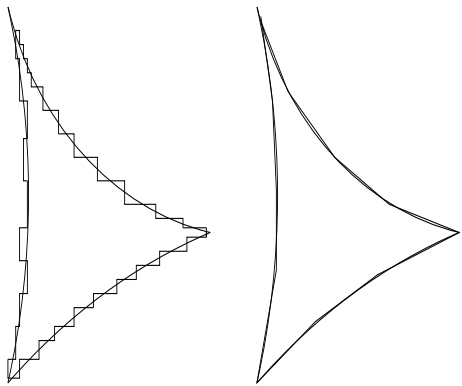,height=5cm,width=6cm,angle=270}}
{\vspace*{.13in}}
\caption[ty]{Two polygonalizations of the 3-disk billiard. We shall
refer to the one on the left as the {\em step-polygon} and the
one on the right as the {\em generalized tangent-polygon} 
since the edges approximate the local tangents. In case the 
polygon is formed from the intersection of local tangents, we
shall refer to it as a {\em tangent polygon} (see fig. \ref{fig:4})
}
\label{fig:1}
\end{figure}

Figure \ref{fig:1} shows two ways of approximating
a 3-disk enclosure by polygons, and, at the classical level, they are
both inequivalent since their invariant surfaces have 
different topologies (see section \ref{sec:classical}). 
Let $\Delta l_{max}$ denote the largest
deviation of the step/tangent-polygon from the smooth billiard
along the boundary. The question that we shall address  
here is : {\em for de Broglie wavelengths, $\lambda > \Delta l_{max}$, 
are these polygons semiclassically equivalent {\rm \cite{equivalent}} 
to the smooth billiard ? }
Naively, one might expect that they are,  since
for $\lambda > \Delta l_{max}$, the system should be unable to distinguish
between the various polygonalized versions and the smooth billiard.
Indeed, such an argument lies at the heart of the discussions in
the work of Cheon and Cohen \cite{cheon_cohen} where they consider
a polygonal version \cite{step} of the Sinai billiard and observe 
GOE (Gaussian Orthogonal Ensemble of random matrices) statistics
in the level fluctuations \cite{dynamics}. However, 
there are various other
instances of polygonal billiards \cite{db_rapid98} which 
exhibit GOE-like fluctuations in a given energy range
but do not resemble any chaotic enclosure. Thus, the
question of semiclassical equivalence cannot be inferred from such
evidence. The work of Tomiya and Yoshinaga \cite{tom_yosh}
is however of greater significance here. They consider polygonalization
of the Bunimovich stadium  by a generalized tangent construction 
\cite{explain} and observe that apart from statistical measures, several
finer features of the stadium are carried over to the tangent polygon.
For instance, a fourier transform of the spectral density reveals a
correspondence between the length spectrum of the stadium and the 
polygon. Besides, individual eigenfunctions in the polygon
exhibit scarring, a phenomenon first observed in the stadium 
billiard \cite{heller_84}. The arguments of Tomiya and Yoshinaga 
however seem to suggest that the only quantities relevant are 
$\lambda$ and $\Delta l_{max}$ so that there might be little
to distinguish between appropriately chosen
tangent and step polygons for $\lambda > \Delta l_{max}$. In
what follows, we shall explore this question in greater detail 
and attempt to provide an answer.

The paper is organized along the following lines. In section 
\ref{sec:classical}, we shall deal with the classical 
aspects of polygonal billiards in general and examine
the special features of tangent polygons.
In section \ref{sec:semiclassics}, we shall first carry out a
semiclassical analysis for generic polygons and
then deal with polygonalized billiards. We shall show 
that generalized tangent-polygons are semiclassically 
equivalent to smooth polygons while step-polygons are not.
A summary of our results and conclusions can be
found in section \ref{sec:conclusions}. 
 
\section{Classical Dynamics }
\label{sec:classical}

\subsection{Polygonal Billiards}

\begin{figure}[tbp]
{\hspace*{0.15cm}\psfig{figure=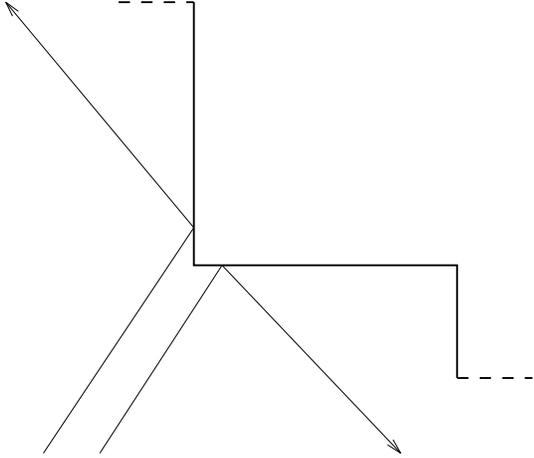,height=6cm,width=7cm,angle=270}}
{\vspace*{.13in}}
\caption[ty]{Parallel rays move away after an encounter with the 
non-integrable ($3\pi/2$) vertex.}
\label{fig:2}
\end{figure}

A notable aspect of generic polygons is the presence of vertices
with internal angles $\{\pi m_i/n_i\}, m_i > 1$. When, $m_i = 1$, the
wedge is {\em integrable} and a parallel band of trajectories continues
to remain parallel after an encounter with the wedge. When $m_i > 1$, 
the band splits up and traverses different paths (see fig.~\ref{fig:2}). 
The presence of many
such vertices leads to multiple splits in the band as it evolves in time.
Interestingly, positive ``effective'' Lyapunov exponents have been
observed in polygonalized billiards \cite{vega}.

The topology of the invariant surface of any polygonal billiard 
can be determined from its genus,

\be
g = 1 + {{\cal N}\over 2} \sum_i {m_i - 1 \over n_i}
\label{eq:genus}
\ee

\noindent
where $\cal N$ is the least common multiple of $\{n_i\}$ \cite{genus}.
Thus, for an integrable
billiard, $g = 1$ so that its invariant surface is topologically
equivalent to a torus. In the context of polygonalized 
billiards, it is interesting to note that while for a circle,
$g = 1$, a polygonalized circle has a very high genus 
\cite{assume_rational}.
Also, the step and tangent-polygons of fig.~\ref{fig:1} have 
different genus while the invariant surface of the smooth 
chaotic billiard is the 3-dimensional constant energy surface. 

In any dynamical system, an important set of trajectories 
are the periodic orbits which live for all times and
close in both position and momentum. In case
of a billiard, the initial and final velocities are
related by a product of reflection matrices, and, for a polygon
where the total number of distinct matrices is finite, it
is possible to obtain conditions for periodicity in
momentum \cite{DB99_2}. We shall use the symbols
$\{1,2,\ldots,N\}$ to denote the $N$ sides 
of the polygon and label trajectories by a 
string of symbols $s_1s_2\ldots s_n$
where $s_i \in \{1,2,\ldots,N\}$. Thus a sequence $1323$
denotes a trajectory that reflects off sides 1, 3, 2 and 3
respectively. Let us denote by $R_i, i=1,N$ the $2 \times 2$ 
reflection matrices of the $N$ sides. These can be 
expressed in terms of the angle $\theta_i$ between the outward normal
to a side and the positive $X$-axis :

\be
R_i = \left ( \matrix{ -\cos(2\theta_i) & -\sin(2\theta_i) \cr
               -\sin(2\theta_i) &\;\;\cos(2\theta_i) } \right ) .
\label{eq:ref_matrix}
\ee 

\noindent
Thus, for the sequence $1323$, the initial and final
velocities are related by

\be
\left ( \begin{array}{c} v_x^f \\ v_y^f \end{array} \right ) = 
R_3 \circ R_2 \circ R_3 \circ R_1 
\left ( \begin{array}{c} v_x^i \\ v_y^i \end{array} \right )
= R_{1323} \left ( \begin{array}{c} v_x^i \\ v_y^i \end{array} \right )
\ee

\noindent
where the superscripts $f$ and $i$ refer respectively to the 
final and initial velocity whose components are $v_x$ and $v_y$.
It is easy to verify that when the number of reflections
is odd

\be
R^{(odd)}_{s_1s_2 \ldots s_n} = 
\left ( \matrix{ -\cos(\varphi_o) & -\sin(\varphi_o) \cr
                 -\sin(\varphi_o) &\;\; \cos(\varphi_o) } \right )
\ee

\noindent
where $\varphi_o = 2 (\theta_1 + \theta_3 + \ldots + \theta_n) -
2(\theta_2 + \theta_4 + \ldots + \theta_{n-1})$ while for 
even number of reflections ($n$ even)

\be
R^{(even)}_{s_1s_2 \ldots s_n} = 
\left ( \matrix{ \;\;\cos(\varphi_e) & \sin(\varphi_e) \cr
                 -\sin(\varphi_e) & \cos(\varphi_e) } \right )
\ee

\noindent
where $\varphi_e = 2 (\theta_1 + \theta_3 + \ldots + \theta_{n-1}) -
2(\theta_2 + \theta_4 + \ldots + \theta_{n})$. 

Obviously, the initial
and final velocities can be equal if the resultant reflection matrix
$R_{s_1s_2\ldots s_n}$ has a unit eigenvalue.
For even $n$ (the case of bands or families), the eigenvalues
are $e^{\pm \imath \varphi_e}$ so that the condition for the
existence of a unit eigenvalue is

\be
\varphi_e = 0 \hspace{0.1in} {\rm mod}(2\pi).
\label{eq:even_condition}
\ee

\noindent
For odd $n$ on the other hand, the product of the eigenvalues
$\lambda_1 \lambda_2 = 1$.
The eigenvector corresponding to a unit
eigenvalue is ($\sin(\varphi_o/2) , -\cos(\varphi_o/2)$) 
so that if a real
orbit exists with the sequence $s_1s_2\ldots s_n$, its initial
and final velocities are equal.

In the event that a sequence repeats itself (denoted by
$\overline{s_1s_2\ldots s_n}$)
and there exists a unit eigenvalue of the resultant matrix
$R_{s_1s_2\ldots s_n}$, stability
considerations guarantee that a periodic orbit exists. 
To see this, consider first an odd bounce orbit ($n$ odd)
for which $s_{n+1} = s_1$ and the initial and final
velocities are equal. Assume further that the initial
and final segments of the trajectory are separated 
by a distance $d$ along the edge, $s_1$. It is then easy to verify
that an isolated periodic orbit exists exactly
in between the two segments (i.e. at a distance $d/2$ from
either segment) with the same velocity.
For the even $n$ case, note first that eq.~(\ref{eq:even_condition})
does not select a particular eigenvelocity. In other words,
there is a range of initial velocities for which (i) $s_1 = s_{n+1}$,
(ii) the trajectories
follows the same sequence $s_1s_2\ldots s_n$  and (iii) the 
initial and final velocities are equal. For convenience,
assume as before that the trajectory starts from edge $s_1$, encounters 
$n$ bounces and  reflects off the same edge after traversing a 
length, $l$, to become
parallel to the initial segment. Further, assume that the two
parallel segments are separated by a distance $d$. It then
follows that a periodic family exists with a velocity 
correction, $\Delta \phi \simeq d/l$. 
In practice, one can rapidly converge to the correct angle
after a few corrections \cite{db_rapid96,db_pramana97}.

We have thus obtained conditions for the existence of periodic
orbits in polygonal enclosures. Note that in the neighbourhood of 
every polygon, $P^{i}$, for which a sequence $s_1s_2\ldots s_n$ 
yields a periodic family, there exists an infinity of polygons
for which this sequence results in a closed almost-periodic (CAP)
family of orbits \cite{DB99_2}. These orbits 
close in position but the angle, $\varphi_e$, between the 
initial and final momentum (at the point where the orbit
closes) is non-zero but small. In contrast, $\varphi_e = 0$
for periodic families. CAP orbit families play a special
role in the classical dynamics of tangent-polygons as we shall
now show.

\subsection{Tangent Polygons}

We shall deal with a (generalized) tangent-polygon 
consisting of $N$ edges which approximates a chaotic 3-disk
billiard as shown in fig.~\ref{fig:1}.
Individual edges thus have an average length $l_{av} = {\cal L}/N$
where $\cal L$ is the perimeter of the smooth billiard.
We shall show here that for $N$ sufficiently large, the
neighbourhood of 
isolated periodic orbits in the smooth billiard 
is well approximated by (i)  closed almost-periodic families of the 
tangent-polygon for even $n$ and (ii) isolated marginally stable periodic
orbits together with the closed orbits in their neighbourhood for odd $n$. 

Consider thus an unstable isolated  periodic orbit in 
the {\em smooth} 3-disk billiard
with symbol sequence $\overline{s_1s_2 \dots s_n}$ where $s_j \in \{1,2,3\}$
corresponding to the 3 sides and $n$ is the number of
bounces or the topological length of the trajectory.
Associated with this periodic orbit is a cylinder of extent
$J_i^{(n)}$ within which all orbits follow the same
symbol sequence \cite{chaos_book,sub_i}. 
Obviously, $J_i^{(n)}$ depends on the
stability and the length of the periodic orbit. In the corresponding
tangent-polygon, if $n$ is reasonably small ($n < n_{max}$)
and $N$ large ($n_{max} = n_{max}(N)$), most trajectories in the 
cylinder survive the sequence in which the polygonalized disks
are visited \cite{same_symbols}.
Note that corresponding to every isolated periodic orbit in the smooth
billiard, there exists a set of $n$ tangents (at the points of impact)
off which the orbit reflects. In the unlikely event that the tangent
polygon has exactly this set as its edges, a periodic orbit trivially
exists in the polygonalized billiard as well. In general
however, the set of tangents can only be {\em approximated} 
by one or more sets of edges in the polygon.

\begin{figure}[tbp]
{\hspace*{0.05cm}\psfig{figure=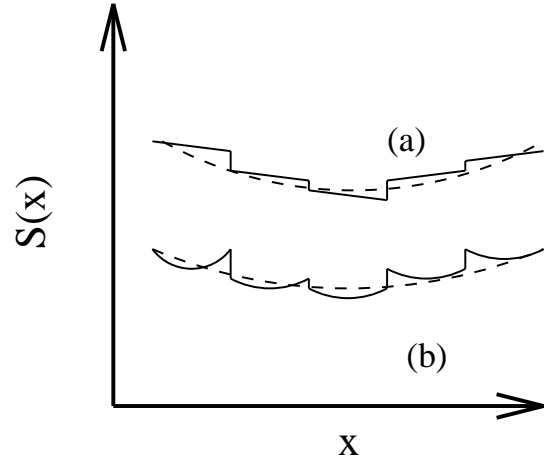,height=6cm,width=7.25cm,angle=0}}
{\vspace*{.13in}}
\caption[ty]{Schematic variation of the action, $S(x)$, in the 
transverse direction, $x = q_{\bot}$ for (a)
even $n$ closed almost-periodic families and (b) odd $n$ closed orbits 
in the tangent polygon. } 
\label{fig:3}
\end{figure}

Consider, first the case when $n$ is even. Clearly, for any 
\cite{more_than_one} set of edges that 
preserves the sequence in which the polygonalized disks 
are visited, $\varphi_e \ne 0$ and hence a periodic orbit 
family does not exist. However, if the approximation
is good, $\varphi_e$ will be small so that a closed
almost-periodic family exists where the angle
between the initial and final momentum is $\varphi_e$.
Within this family, the difference in length of two orbits
is $\varphi_e q_\bot$ where $q_\bot$ is the
transverse separation between the two \cite{DB99_2}. 
Thus, if $\varphi_e$ is
small, the variation in length within the family is slow.
In general, there can be more than a single set of such 
closed almost-periodic family depending on $N$ and the
stability of the isolated periodic orbit
and for one of these, the average
length will be close to that of the isolated periodic
orbit. The variation in length of closed almost-periodic orbits
is schematically shown in fig.~\ref{fig:3} (case a)
where the dashed curve describes the neighbourhood 
of the isolated periodic orbit. Clearly, with an 
increase in $N$, the number of families increase,
their widths decrease
and the approximation of the neighbourhood by closed
almost-periodic families gets better.

For the case of odd $n$, assume that an isolated 
unstable periodic orbit exists for the sequence $s_1s_2\ldots s_n$
in the smooth billiard with  an initial velocity that can 
be calculated from the set of tangents to the boundary
at the points of impact. For every set (of edges)
that approximates these tangents, there exists in the polygon an isolated 
marginally stable periodic orbit with a slightly 
different initial velocity owing to the difference
in $\varphi_o$. On either side however, 
there exist closed orbits that follow
the same sequence and whose length increases  as
$l(q_\bot) = l(0) \left [ 1 + (2q_\bot/l(0))^2 \right ]^{1/2}$
where $l(0)$ is the length of the periodic orbit and $q_\bot$
measures the transverse distance from it. 
As before, for every set of edges that approximates
the tangents, one can observe this behaviour so that the  
variation in length in the neighbourhood of the unstable isolated
periodic orbit is schematically as shown in fig.~\ref{fig:3} (case (b)).

Apart from the closed almost-periodic orbits, there exist an 
infinite number of exact periodic orbit families in every 
polygonalized billiard. The extent of these families is limited
by the length of the smallest edge and within each such family,
periodic orbits have identical length. For $N$ sufficiently large
however, along with every set of edges that gives rise to an 
exact periodic family, there exist other sets of edges for 
which orbits follow the same sequence but are almost-periodic.
The variation in length is thus only marginally different 
from that shown schematically in fig.~\ref{fig:3} (case a) with 
one band having constant length while in the others, the 
length changes linearly within the band.

\subsection{Step Polygons etc.}

We now turn our attention to the step polygon of fig.~\ref{fig:1}.
Obviously, the edges do not locally approximate the tangent
at any point of the smooth billiard so that the cylinder $J_i^{(n)}$
does not survive even for small $n$. In fact, there exist several
families of bouncing ball orbits in the step polygon of fig.~\ref{fig:1}
which have no correspondence in the smooth billiard.
Besides, all periodic orbits have even $n$ and closed
almost-periodic families do not exist in this case.
Thus the classical dynamics in the neighbourhood of isolated
unstable periodic orbits is not approximated by closed orbits
in the step polygon.

There are several other methods of polygonalizing smooth billiards
and in each of these, the symbolic dynamics of the smooth billiard
cannot survive unless the edges locally approximate the tangents.
In summary then, generalized tangent polygons are the only
systems in which the classical dynamics of smooth billiards is
locally preserved for $n < n_{max}$.

\section{Semiclassics}
\label{sec:semiclassics}

\subsection{Generic Polygons}

Having established a correspondence between the classical dynamics of the
smooth billiard and  the tangent polygon, we shall now 
consider the quantum problem and
derive a semiclassical expression for its density of states.
The starting point in such an analysis is the relation \cite{gutzwiller}

\bea
\sum_n {1\over E - E_n} & = & \int~dq~G(q, q; E) \\ 
                        & \simeq & \int~dq~G_{s.c.}(q, q; E) \label{eq:basic}
\eea

\noindent
where $\{E_n\}$ are the energy eigenvalues and 
$G_{s.c.}$ refers to the semiclassical energy dependent 
propagator (Green's function). For a polygonal billiard,  

\be
G_{s.c}(q, q'; E) = - \imath \sum {1\over \sqrt{8\pi \imath k l(q,q')}}
e^{\imath k l(q,q')  - \imath \mu \pi/2} \label{eq:G_sem}
\ee

\noindent
where the sum runs over all orbits at energy $E = k^2$ between $q$ and $q'$  
having length $l(q,q')$ and $\mu$ is twice the number ($n$) of 
reflections at the boundary. For convenience,
we have chosen the mass $m = 1/2$ and $\hbar = 1$

In the limit $k \rightarrow \infty$, 
the only trajectories that survive the 
trace operation are the periodic orbits 
\cite{gutzwiller}. As mentioned earlier, even bounce periodic orbits
occur in families over which the length of the orbit does not vary
and for these $\int dq = a_p$ where $a_p$ refers to the area 
occupied by a primitive periodic orbit \cite{pj_mv}. 
For the odd-bounce case, a local co-ordinate system ($q_{\|},q_{\bot}$)
needs to be introduced where $q_{\|}$ is the position along the 
isolated periodic orbit and $q_{\bot}$ measures the transverse 
distance from the periodic orbit. Since the length does not
vary along the orbit, $\oint q_{\|} = l_{p'}$ where $l_{p'}$ is
the length of the primitive periodic orbit. The $q_\bot$ 
integration can be performed by the stationary phase approximation
using the expression for $l(q_\bot)$ given earlier.
Thus,

\bea
\rho(E) & = & \sum_n \delta (E - E_n) = - {1\over \pi} \lim_{\epsilon
\rightarrow 0} \Im {1\over E + i \epsilon - E_n} \nonumber \\
        & \simeq & \rho_{av}(E) + \sum_p \sum_{r=1}^{\infty} {a_p \over
\sqrt{8\pi^3 k rl_p}} \nonumber \\
& \times & \cos(k rl_p - \pi/4) -
\sum_{p'} \sum_{r'=1}^{\infty} {l_{p'} \over 4\pi k} \cos(k r'l_{p'})
 \label{eq:semi_poly}.
\eea    

\noindent
where $\rho_{av}$ is the average density of states and
the sums over $p$ and $p'$ run over (primitive) {\em families}
and isolated orbits respectively having length $l_p$ and $l_{p'}$.

For finite $k$ however, eq.~(\ref{eq:semi_poly}) is inadequate
for generic polygons and the most prominent correction that has 
so far been taken into account arises from diffraction 
\cite{pavloff,bruss_whelan,sieber_etal}. 
It has recently been shown \cite{DB99_2} 
that closed almost-periodic orbit families contribute as well
and  with weights comparable to those of periodic orbit families
when $\varphi_e$ is small. The correct trace formula 
can be derived by noting that for a closed almost-periodic
family, $l(q_\bot) = l(0) + q_\bot \varphi_e$
where $l(0) = l_i$ is the length of the orbit in 
the centre of the band and $q_\bot$ varies from $-w_p/2$ to
$w_p/2$. Assuming that $k$ is sufficiently large, the amplitude
($1/l(q_\bot)$) can be treated as a constant ($1/l_i$)
and the trace formula for finite $k$ is then 

\bea 
\rho(E) & \simeq & \rho_{av}(E) + \sum_i  {a_i \over
\sqrt{8\pi^3 k l_i}} \nonumber \\
& \times & \cos(kl_i  -  \pi/4) 
{\sin(k\varphi_e^{(i)} w_i/2)\over k\varphi_e^{(i)} w_i/2}
 \nonumber \\ 
& - & \sum_{p'} \sum_{r'=1}^{\infty} {l_{p'} \over 4\pi k} \cos(k r'l_{p'}).
\label{eq:semi_poly_modif}
\eea

\noindent
In eq.~(\ref{eq:semi_poly_modif}),
 the sum over $i$ runs over closed almost-periodic and periodic
orbit families and $l_i$ is the (average) length
of such a family. Note that as $k \rightarrow \infty$, the 
contribution of CAP families  ($\varphi_e^{(i)} \ne 0$)
vanishes and eq.~(\ref{eq:semi_poly_modif}) reduces to 
eq.~(\ref{eq:semi_poly}). For de Broglie wavelength, $\lambda
 >> \pi (w_p \varphi_e^{(i)})$, however,
the ($i$th) closed almost-periodic orbit family contributes
with a weight comparable to that of periodic families (${\cal O}(1/k^{1/2})$)
and hence assumes greater significance than diffraction.
Interestingly, such orbits clearly show up in eigenfunctions
\cite{bellomo} and the phenomenon has been referred to as 
``scarring by ghosts of periodic orbits'' since such a
periodic orbit exists only in a neighbouring polygon.

Note that almost-periodic closed orbit families do not generally occur 
in systems where the number of directions accessible to a trajectory 
is small since the (average) angle of intersection is large. Thus 
in step billiards or in systems with very low genus, almost-periodic
families do not contribute significantly. In generic polygons however,
these orbits are the key to semiclassical quantization. 

\subsection{Tangent Polygons}

Though eq.~(\ref{eq:semi_poly_modif}) holds for generic polygons,
we shall   use a somewhat different approach for the tangent-polygons
where several families of closed almost-periodic families
exist together with a gradual change in length. This variation
can be approximated by a smooth curve as depicted schematically
in fig.~\ref{fig:3}. For $N$ sufficiently large,
it is reasonable to choose the smooth curve as the one which describes
the (linearized) neighbourhood of the isolated unstable periodic
orbit in the chaotic billiard and the error so introduced can
be evaluated.  Thus instead of 
summing over nearby bands we shall carry out a single
integration for every periodic sequence. The trace operation
then leads to the truncated Gutzwiller trace formula

\bea
\rho(E) & \simeq & \rho_{av}(E)  +  {1\over k} [
\sum_{T_p < T^*} \sum_{r=1}^{\infty}
 {l_p \over 2\pi \sqrt{\left |\det(J_p^r - I) \right |} } \nonumber \\      
  & \times & \cos({k rl_p} - r\mu_p \pi/2)
 + \Delta\rho_1 + \Delta\rho_2 ]
\label{eq:gutz}
\eea 

\noindent
with errors $\Delta\rho_1$ and $ \Delta\rho_2$. In eq.~(\ref{eq:gutz}),
$T_p = l_p/2k$ is the time period of a primitive periodic orbit, $J_p$ is
the Jacobian matrix arising from a linearization of the flow 
in the neighbourhood of a periodic orbit and $\mu_p$ is the 
Maslov index associated with the primitive orbit.

Of the errors, the first, $\Delta\rho_1$, 
arises due to the restriction of the periodic orbit sum
to orbits of length $T_p < T^*$ since the correspondence 
between the smooth billiard and the tangent polygon exists
only for $n < n_{max}$. Obviously, $\lim_{N \rightarrow \infty} 
\Delta\rho_1 = 0 $. We shall however consider 
tangent polygons for which $T^* > T_H$ ($T_H$ is the
Heisenberg time) such that the energy eigenvalues 
evaluated using the truncated periodic orbit sum gives
a faithful approximation to the true eigenvalues $\{E_n\}$.
Thus, we shall neglect $\Delta\rho_1$ henceforth. 

The error $\Delta\rho_2$ arises due to the approximation
shown in fig. \ref{fig:3} and a crude 
estimate of this can be obtained by assuming that the 
length is constant within a band. The error 
is then

\be
\Delta\rho_2 \simeq \sum_i {w_i^2 \over 2} f'(x_i)
\ee

\noindent
where $w_i$ is the width of the $i$th band, $x_i$ is the 
value of $q_\bot$ within the $i$th band for which $l(q_\bot)$
equals the value of the smooth curve and

\be
f(q_\bot) \sim {k^{1/2}} e^{{\imath} k l(q_\bot)}
\ee

\noindent
Thus, 

\be
\Delta\rho_2   \sim  \sum_i w_i^2 k^{3/2}
	       \sim  l_{av}^2 k^{3/2} 
	       =  { {\cal L}^2 k^{3/2} \over N^2}	
\ee

\noindent
In writing the above we have used the facts that the maximum width, 
$w_i$ of a periodic orbit band is limited by the average length, $l_{av}$
of the edges. We may further assume that the number of families
corresponding to a cylinder $\overline J_i^{(n)}$ is
small compared to $N$ so that $\Delta\rho_2 \sim k^{3/2}/N^2$.
Thus for $k << {\cal C} N^{4/3}$, the tangent polygon is semiclassically
equivalent to the smooth billiard where ${\cal C}$ is a positive 
constant which depends on the exact form of $f(x)$.

It may be noted that a special construction of tangent polygon 
occurs in the boundary integral method of evaluating quantum
eigenvalues. Here, the Schroedinger equation is reduced
to an eigenvalue problem for an integral operator $K$ \cite{boasmann}

\bea
\psi(s) &  = & \oint ds' \psi(s') K(s,s';k) \label{eq:bim_main} \\
K(s,s';k) & = & - {\imath k \over 2} \cos \theta(s,s') H_1^{(1)}
(k|\vec{s} - \vec{s'}|) \label{eq:bim_1} \\
\cos \theta(s,s') & = & \hat{n}(\vec{s}).\hat{\rho}(s,s')
\label{eq:bim_2}
\eea

\noindent
where $E = k^2$, $\hat{\rho}(s,s') =
(\vec{s} - \vec{s'})/|\vec{s} - \vec{s'}|$
and $ \hat{n}(\vec{s})$ is the outward normal at the point $\vec{s}$.
The unknown function is now the normal derivative
on the boundary
\be
\psi(s) = \hat{n}(\vec{s}).\nabla \Psi(\vec{s})
\label{eq:psi_Psi}
\ee

\noindent
and the full interior eigenfunction can be recovered through the
mapping   

\be
\Psi(q) = - {\imath \over 4} \oint ds H_0^{(1)}(k|\vec{s} - \vec{s'}|) \psi(s).
\label{eq:Psi_psi}
\ee
   
\noindent
In practice, the boundary is discretized with the number of
points $N \simeq {\cal L} k/ \pi$ \cite{bogo}. Eq.~(\ref{eq:bim_main})
then reduces to a matrix equation leading to the consistency
condition 

\be
\det(I - \Delta l~K(k)) = 0
\ee

\noindent
where $\Delta l$ is the incremental distance along the boundary
and $I$ is the identity matrix.
Note that for a straight edge, $K_{nn} = 0$ while for
a curved boundary, $K_{nn} = \pm 1/(2\pi R_n)$
where $R_n$ is the local radius of curvature at the
boundary point $s_n$ and the $+$ and $-$ signs are
for convex and concave boundaries respectively.

\begin{figure}[tbp]
{\hspace*{0.5cm}\psfig{figure=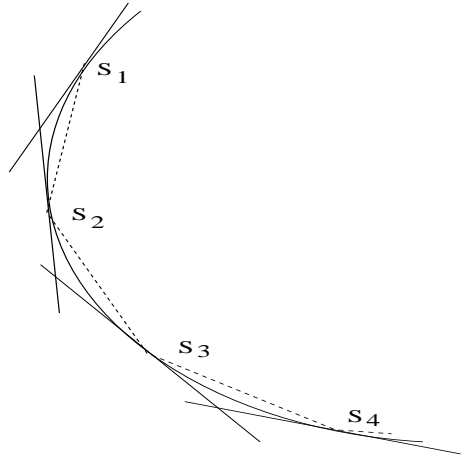,height=6cm,width=6cm,angle=270}}
{\vspace*{.13in}}
\caption[ty]{Construction of a polygon using the tangents at 
the points $\{s_n\}$.}
\label{fig:4}
\end{figure}

The corresponding tangent polygon may be constructed by the intersection
of the tangents at the boundary points $s_n$ as shown
in fig.~\ref{fig:4} and this in turn may be solved

\noindent
using the boundary integration method with the same
set of points $\{s_n\}$. The only difference then would
be the local curvature in the diagonal matrix element, 
which for any polygon is zero.
The error so generated is similar to the approximation
of fig.~(\ref{fig:3}) where we replace the steps
by a smooth curve which contains information about the
local curvatures at the points of impact.

To test this assertion, we have evaluated the eigenvalues
using the boundary integral method for an intersecting 3-disk
system with (a) $K_{nn} = 0$ in one case and (b) $K_{nn} 
= 1/(2\pi R_n)$ in the other. In both cases, $N = 2{\cal L} k/\pi$ 
where ${\cal L} = 1$. Table~\ref{table:1} lists the 
four sets of eigenvalues in the range $1505 < k < 1506$.
The first two are the exact quantum eigenvalues for
the 3-disk and the tangent-polygon ($K_{nn} = 0$) while
the eigenvalues in the third and fourth columns are
determined using the asymptotic form of the Hankel
function, $H_1^{(1)}$ and are referred to as the
``semiclassical'' eigenvalues \cite{bogo}. 
Clearly, the polygonalization error is small compared to
the semiclassical error so that the two systems are equivalent.

\subsection{Circles , Step Polygons etc.}

It is easy to see that the analysis carried out so far holds 
for other smooth billiards which are non-chaotic and where periodic
orbits may occur in families. An extreme case is the circle billiard
where under similar conditions, a one to one correspondence between 
its periodic orbits and those of the tangent polygon exists.
However, rather than a single family of periodic orbits with 
a sharply defined action and angular momentum, there exists in the 
tangent polygon, a number of closed almost-periodic bands
(or isolated periodic orbits and the associated closed
orbits in its neighbourhood when $n$ is odd). When the 
variation in length across these families is small ($N$ large),
it can be replaced by a constant length typical of periodic
orbit families and the error so generated can be similarly
evaluated.
  
In case of the step polygon of fig.~\ref{fig:1}, only
the Weyl term in the density of states agrees with that of the
3-disk billiard as the areas can be made identical in an 
appropriate construction. The contributions from periodic 
orbits are bound to differ as explained in section \ref{sec:classical}
and diffractive corrections are both significant and 
different from those in the 3-disk. Thus, the two systems 
are inequivalent.

\section{Conclusions}
\label{sec:conclusions}

We have addressed the question of semiclassical equivalence
of polygonalized billiards in this paper and in the process analyzed
the conditions under which orbits are periodic in generic polygons.
We have also provided a trace formula for finite energy that includes 
contributions from closed almost-periodic orbit families. Since
their weights can be comparable to those of periodic orbit families,
such orbits must be included in any realistic semiclassical
calculation.

In summary, polygonalized billiards are {\em semiclassically} 
equivalent to smooth billiards in appropriate energy
ranges only when 
the edges locally approximate the tangents to the boundary of
the smooth billiard. In other cases such as the step polygon, 
the classical dynamics has no correspondence with the smooth billiard
and the two are not semiclassically equivalent. 

The results of this paper can be applied to statistics of quantum
energy levels with interesting consequences and we shall briefly
discuss these here. It is obvious that given any smooth 
billiard, there exist tangent polygons whose energy levels
faithfully approximate those of the smooth billiard in a range
that increases with the number of sides in the polygon.  
Thus, the level statistics in this range can vary from Poisson
to GOE depending on the statistics of the smooth billiard.
At finite energies therefore, polygonal billiards do not 
belong to any universality class. This however does not
preclude the existence of universality in a sub-class of
polygons such as generic triangular billiards. It also follows that
level statistics at finite energies does not depend on the genus within
the broad class of polygons.

\newcommand{\PR}[1]{{Phys.\ Rep.}\/ {\bf #1}}
\newcommand{\PRL}[1]{{Phys.\ Rev.\ Lett.}\/ {\bf #1}}
\newcommand{\PRA}[1]{{Phys.\ Rev.\ A}\/ {\bf #1}}
\newcommand{\PRB}[1]{{Phys.\ Rev.\ B}\/ {\bf #1}}
\newcommand{\PRD}[1]{{Phys.\ Rev.\ D}\/ {\bf #1}}
\newcommand{\PRE}[1]{{Phys.\ Rev.\ E}\/ {\bf #1}}
\newcommand{\JPA}[1]{{J.\ Phys.\ A}\/ {\bf #1}}
\newcommand{\JPB}[1]{{J.\ Phys.\ B}\/ {\bf #1}}
\newcommand{\JCP}[1]{{J.\ Chem.\ Phys.}\/ {\bf #1}}
\newcommand{\JPC}[1]{{J.\ Phys.\ Chem.}\/ {\bf #1}}
\newcommand{\JMP}[1]{{J.\ Math.\ Phys.}\/ {\bf #1}}
\newcommand{\JSP}[1]{{J.\ Stat.\ Phys.}\/ {\bf #1}}
\newcommand{\AP}[1]{{Ann.\ Phys.}\/ {\bf #1}}
\newcommand{\PLB}[1]{{Phys.\ Lett.\ B}\/ {\bf #1}}
\newcommand{\PLA}[1]{{Phys.\ Lett.\ A}\/ {\bf #1}}
\newcommand{\PD}[1]{{Physica D}\/ {\bf #1}}
\newcommand{\NPB}[1]{{Nucl.\ Phys.\ B}\/ {\bf #1}}
\newcommand{\INCB}[1]{{Il Nuov.\ Cim.\ B}\/ {\bf #1}}
\newcommand{\JETP}[1]{{Sov.\ Phys.\ JETP}\/ {\bf #1}}
\newcommand{\JETPL}[1]{{JETP Lett.\ }\/ {\bf #1}}
\newcommand{\RMS}[1]{{Russ.\ Math.\ Surv.}\/ {\bf #1}}
\newcommand{\USSR}[1]{{Math.\ USSR.\ Sb.}\/ {\bf #1}}
\newcommand{\PST}[1]{{Phys.\ Scripta T}\/ {\bf #1}}
\newcommand{\CM}[1]{{Cont.\ Math.}\/ {\bf #1}}
\newcommand{\JMPA}[1]{{J.\ Math.\ Pure Appl.}\/ {\bf #1}}
\newcommand{\CMP}[1]{{Comm.\ Math.\ Phys.}\/ {\bf #1}}
\newcommand{\PRS}[1]{{Proc.\ R.\ Soc. Lond.\ A}\/ {\bf #1}}

\vskip 0.25 in

\begin{table}[tbp]
\begin{tabular}{llll}
$k_{3disk}^{exact}$ & $k_{polygon}^{exact}$ & $k_{3disk}^{semi}$ & 
$k_{polygon}^{semi}$ \\\hline
1505.11325 & 1505.11324 & 1505.16810 & 1505.16809 \\  
1505.17526 & 1505.17527 & 1505.24944 & 1505.24945 \\
1505.37518 & 1505.37518 & 1505.43624 & 1505.43624 \\
1505.49791 & 1505.49792 & 1505.55890 & 1505.55890 \\
1505.59969 & 1505.59969 & 1505.65119 & 1505.65119 \\
1505.72383 & 1505.72382 & 1505.78927 & 1505.78926 \\
1505.77798 & 1505.77798 & 1505.83193 & 1505.83192 \\
1505.83710 & 1505.83709 & 1505.89083 & 1505.89082 \\
\end{tabular}
\vskip 0.1 in
\caption{A comparison of the exact and semiclassical eigenvalues
of the 3-disk with the exact and semiclassical eigenvalues of
the corresponding tangent polygon. Neglect of the curvature
term makes little difference compared to the semiclassical
approximation.
\label{table:1}}
\end{table}


\begin{references}
\bibitem{equivalent} In this paper, two billiards are said to  
be semiclassically equivalent if their leading order
contributions to the semiclassical trace
formula are identical. The semiclassically quantized energy levels
are thus identical.
\bibitem{step} The circle is approximated by steps in this case.
\bibitem{dynamics} The quantum energy eigenvalues are known
to obey $GOE$ statistics when the corresponding classical system 
is chaotic \cite{bgs_84}.
\bibitem{explain} In order to preserve area, the edges that
replace the semi-circles are not tangents. We shall 
use the term ``generalized tangent-polygon'' in all cases where 
the edges approximate the local tangents to the smooth billiard.
\bibitem{genus} A surface with genus, $g$, is topologically equivalent
to a sphere with $g$ holes.
\bibitem{more_than_one} There can be more than a single set
of edges for which the symbol sequence is the same 
depending on $N$ and the stability of the periodic orbit in the
smooth billiard.
\end{references}
\end{document}